# Resonances On-Demand for Plasmonic Nano-Particles

Pavel Ginzburg[*], Itay Shor, Nikolai Berkovitch, Amir Nevet, and Meir Orenstein

*EE department, Technion – Israel Institute of Technology, Technion City, Haifa 32000 Israel*

*\*gpasha@tx.technion.ac.il*

**Abstract:** A method for designing plasmonic particles with desired resonance spectra is presented. The method is based on repetitive perturbations of an initial particle shape while calculating the eigenvalues of the various quasistatic resonances. The method is rigorously proved, assuring a solution exists for any required spectral resonance location. Resonances spanning the visible and the near-infrared regimes, as designed by our method, are verified using finite-difference time-domain simulations. A novel family of particles with collocated dipole-quadrupole resonances is designed, demonstrating the unique power of the method. Such on-demand engineering enables strict realization of nano-antennas and metamaterials for various applications requiring specific spectral functions.

Localized resonances of plasmonic nano-particles, stemming from the unique interplay between local geometry and dielectric dispersion, may be directed towards novel applications. Recent technological progress enabled the control of a particle shape with nanometric accuracy and opened up new possibilities for practical and prospective applications, including: enhanced sensing and spectroscopy [1], plasmonic biosensors [2], cancer imaging and therapy [3], building block of metamaterials [4,5,6], plasmonic lasers [7,8] spasers [9], enhanced nonlinearities [10,11], enhancement of radiation efficiencies [12], etc.

Typical metallic nanoparticle geometries, e.g. spheres, disks [13], and bow-ties [14] placed on low index dielectric substrates, exhibit resonances at wavelengths primarily below 1µm. The resonance extension to the near-infrared (NIR) regime (1–2µm) is of great importance for optical communications and biomedical applications. Until today this was accomplished by coupling several particles, resulting in hybridization of the resonant modes and modification of the resonances [15]. Recently, new types of particles resonating at NIR were demonstrated [16] emphasizing the role of local geometry, and more specifically of concave geometry.

Although the task of designing nanoparticles with predetermined resonance wavelengths is of great interest, no generic method was hitherto presented. The existing methods for resonance engineering rely on parametric variation of certain dimension of a specific particle [17], or inter-particle coupling distance [18,19], but may be constrained by technological limitations. While evolutionary algorithms of different kinds [20,21] are widely used in electromagnetic research for optimization on excitation signal for a given shape [22,23] and for antenna engineering [24,25,26], they were not implemented in the field of nano-optics, to the best of our knowledge. Moreover, most of the existing algorithms are heuristical and, in principal, may not converge to an optimal solution.

Here we propose and demonstrate on-demand engineering of the multipole plasmon resonances of a sub-wavelength particle. The method is based on a series of small perturbations applied to an (arbitrary) initial particle, which enables the modification of the spectral location of a resonance or multiple resonances according to specific predetermined values. Moreover, resonances of different multipole orders are designed to be degenerated at specific wavelengths, e.g. the dipole and quadrupole resonances are collocated at the same wavelength. The resonances and the field distribution calculated analytically by our method are subsequently verified using forward scattering finite difference time domain (FDTD) simulations.

Small perturbations of an initial particle geometry yield small shifts of its resonances. Here we show that a proper sequence of perturbations is capable of shifting the resonances towards the requested values. Since a resonance is determined by the material permittivity $\varepsilon(\omega)$ at the resonance frequency $\omega$, the problem may be formulated in terms of $\varepsilon(\omega)$ rather than $\omega$. Suppose that a pair of initial resonances $\varepsilon_1, \varepsilon_2$ of a polygonal shaped nano-particle should be shifted towards a new pair. The following perturbation will be applied – each side $k$ of the initial polygon will be associated with a real number $a_k$. The side will then be moved in the direction of the local normal by a distance of $h \cdot a_k$, where $h$ is a small dimensionless parameter, common to all sides. The formalism of resonance shifting due to such shape perturbation of a plasmonic particle was developed in [27] resulting in:

$$\frac{\partial \varepsilon}{\partial h} = (1-\varepsilon) \frac{\int_{\partial \Omega} d^2 S a \left( \left(\nabla_\partial u_{in}\right)^2 + \varepsilon \left(\partial_n u_{in}\right)^2 \right)}{\int_{\partial \Omega} d^2 S u_{in} \partial_n u_{in}} \tag{1}$$

where $\frac{\partial \varepsilon}{\partial h}$ is the derivative of the dielectric permittivity in respect to the dimensionless parameter $h$, $u_{in}$ the electric potential inside the unperturbed particle, $\partial \Omega$ the particle boundary, $\partial_n$ a derivative normal to the boundary direction, $\nabla_\partial$ a derivative tangent to the boundary direction, and $a$ a scalar function along the boundary, indicating the amount of particle deformation in the direction of the local normal. Suppose that we move only two arbitrary sides on the discretized boundary by a fraction of $h \cdot a_1$ and $h \cdot a_2$ correspondingly, and consider the following measure $S_{1,2} = \frac{\partial \varepsilon_{1,2}/\partial h}{(1-\varepsilon_{1,2})} \int_{\partial \Omega} d^2 S u_{in} \partial_n u_{in}$. Only the sign of $S_{1,2}$ is of importance, since it indicates the direction of the eigenvalue modification after the applied variation. If any combination of signs is possible, it means that any pair of initial resonances may be shifted towards any independent location. Eq. 1 may then be rewritten as:

$$\begin{pmatrix} S_1 \\ S_2 \end{pmatrix} = \begin{pmatrix} \mu_1 + \varepsilon_1 \eta_1 & \mu_2 + \varepsilon_1 \eta_2 \\ \mu_1 + \varepsilon_2 \eta_1 & \mu_2 + \varepsilon_2 \eta_2 \end{pmatrix} \begin{pmatrix} a_1 \\ a_2 \end{pmatrix}, \quad \mu_{1,2} = \left(\nabla_\partial u_{in}(x_{1,2})\right)^2, \eta_{1,2} = \left(\partial_n u_{in}(x_{1,2})\right)^2 \quad (2)$$

This equation is solvable for nondegenerate resonances, since the determinant of the matrix is not zero. Consequently, since the rank of the relevant geometrical transformation is 2, up to two nondegenerate eigenvalues (multipole resonances) may be moved towards any requested values. Specifically, any single resonance (dipole, quadrupole, etc) may be shifted as required.

For a given arbitrary shape (polygon) all the resonances may be calculated by solving the 'direct problem' of extracting the spectral eigen value for the resonances of a known geometry and material parameters. Here we used discretized boundary integral method, yielding a matrix equation for the eigen values (function of dielectric permittivity) and eigen vectors (surface charge distribution) [e.g. 28]. Any other solver may be employed including FDTD [29], finite element method, and volume integral method. In order to reduce the calculation complexity, we formulate the problem in two dimensions – infinite cylinder with polygonal cross section. Small perturbations to a particle geometry yield small shifts of its resonances, which are tracked by the 'direct solver'. Each applied perturbation requires verification of the shape validity according to topological and technological limitations: minimal angles (to avoid sharp corners), minimal width (to avoid high aspect ratios), and no-intersections. The perturbation is applied in the following way – two opposite sides of a symmetric shape are moved by a fraction $\delta$ in the direction of the local normal tilted by an angle $\theta$ (preserving the initial symmetry). $\theta$ is an additional degree of freedom used for faster convergence. The particle is subsequently smoothed to prevent corners. If the resulting shape satisfies the desired properties then all resonances are recalculated. The perturbation will be accepted if the resonances will move towards their required values otherwise the perturbation is nullified and the algorithm step is repeated.

By applying this method we produced a series of particles with resonances spanning the entire visible and NIR spectrum and verified their transmission spectra by FDTD simulations using a proper excitation. These explicit examples show that the entire spectrum may be optimized for absorption and serve, for example, as a building block in solar cells [30] and related applications. The transmission spectrum (forward scattering) of each particle is presented in Fig. 1, where the insets are the corresponding particle geometries and arrows correspond to the proper polarization of their excitation. This numerical experiment also corresponds to normal incidence spectroscopic measurements, using an array of particles rather than a single particle, in order to enhance the transmitted signal [16]. It should be noted that quality factors of plasmonic resonances, evaluated in our calculations to be ~20, are usually limited to values below 100 and are higher at the visible spectrum than at NIR [31]. However, the usefulness of the localized plasmon resonances does not emerge from their quality factor, but rather from their associated small modal volumes. The simulated particle transversal dimensions were taken to be 100nm in diameter (maximal distance between two points on the object); however, this parameter was found to have minor influence, since all the resonances have quasi-static characteristics. The particle material was chosen to be Au (including losses) with appropriate dispersion function [32]. Each transmission dip within the spectrum corresponds to an engineered dipole resonance and fits the values predicted by our method within ~2% accuracy. The resonances of convex gold particles with small aspect ratios are situated in the visible part of the spectrum; hence the shifting to longer wavelengths increases the number of required perturbations, as may be seen from the final shapes (Fig. 1). Fabrication of such particles, resonant at NIR, is definitely feasible with today's mature nano-technology, although it may require advanced lithography such as E-beam lithography, including a correction of the proximity effect (interaction of electron beam with resist and substrate). All these significant issues were constrained in the algorithm.

Although we demonstrate the algorithm for 2D particles in order to reduce the calculation time, the presented method may provide shapes in any dimension, in particular in 3D, and is therefore a very powerful tool. Moreover, particles with cylindrical cross-section and finite height are of potential interest for applications and they may be related directly to 2D shapes with phenomenological Lorentz depolarization factor [33,34]. According to this model - increasing a particle dimension in the direction of the excitation polarization shifts the particle resonance to the red, while if such an increase is made in the perpendicular direction, the particle's resonance exhibits a blue shift. We verified it by performing forward scattering simulations for particles of 50, 100 and 500nm height (Fig. 2). Very similar linear behavior of the resonance location was observed for all cases, exhibiting a red shift from the infinitely thick (2D) to the thin (50nm) particle, demonstrating that the resonance location for this 3D scenario can be deduced to a good approximation from the 2D calculation.

To further exploit the capabilities of our method we generated a new family of particles in which two resonances of different orders are brought to be frequency degenerated at a desired wavelength. An initially 'cross-shaped' particle (left inset of Fig. 3) with dipole and quadrupole resonances separated by about 30nm (in 2D geometry), is mutated to a particle (right inset of Fig. 3) with the dipole (Fig.4 (a)) and quadrupole (Fig.4 (b)) resonances collocated at 600nm. Particle parameters are similar to those of Fig. 1. This unique phenomenon may contribute to applications such as optimal nano-scale absorber [35], emitters' lifetime manipulation, and near field storage devices. To examine the unique behavior of such particles, we performed FDTD simulations, in which a monochromatic dipole was placed in the vicinity of each of the two particles: the perfect cross-shaped, and the above discussed 'degenerate-resonance' particle (insets of Fig. 3). The overall near-field intensity in the vicinity of each particle was calculated as a function of wavelength (Fig.3) and normalized by the results obtained at free space (no particle). Since the mutated particle is degenerated in frequency, the two resonances (dipole and quadrupole) were simultaneously coupled to the excitation, creating a coherent superposition of two modes. For perfectly separated resonances of the initial particle, this degenerated combination should result in ~4-fold intensity enhancement (assuming equal excitation efficiency) in comparison to the standalone resonance, while in this specific case the factor of 3 is obtained due to a partial overlap of finite-width resonances of the cross-shaped particle. The intensity enhancement for the degenerate resonance particle, relative to a free space, is 45.

In conclusion, we demonstrated a generic method for the engineering of plasmon resonances. The algorithm is capable of producing particles with desired resonances over the entire spectrum. Moreover, unique collocation of several resonances may be achieved and employed for storage of the near field energy and for absorption enhancement.

**Figures Captions:**

**Figure 1.** Transmission spectra of engineered particles. Insets – respective particles shapes. 'A' particle – blue dashed line, 'B' – red circles, 'C' – green dash-dot, 'D' – black solid.

**Figure 2.** Resonances for 3D particles - infinite height (black triangle), 500nm (green circle), 100nm (blue diamond) and 50nm (red rectangle) thick disk particles. Particles numbers '1'-'4' correspond to 'A'–'D' particles from Fig. 1.

**Figure 3**. Intensity enhancement of a cross-shpaed particle (black line) and a 'degenerate' particle (red line) (normalized to the free space scenario). Insets – corresponding particles and their excitation by a radiating dipole.

**Figure 4.** A particle with collocated resonances and the corresponding electrical fields. Color on the boundary indicates the surface charge distribution. (a) dipole, (b) quadrupole.

**List of figures:**
**Figure 1.**

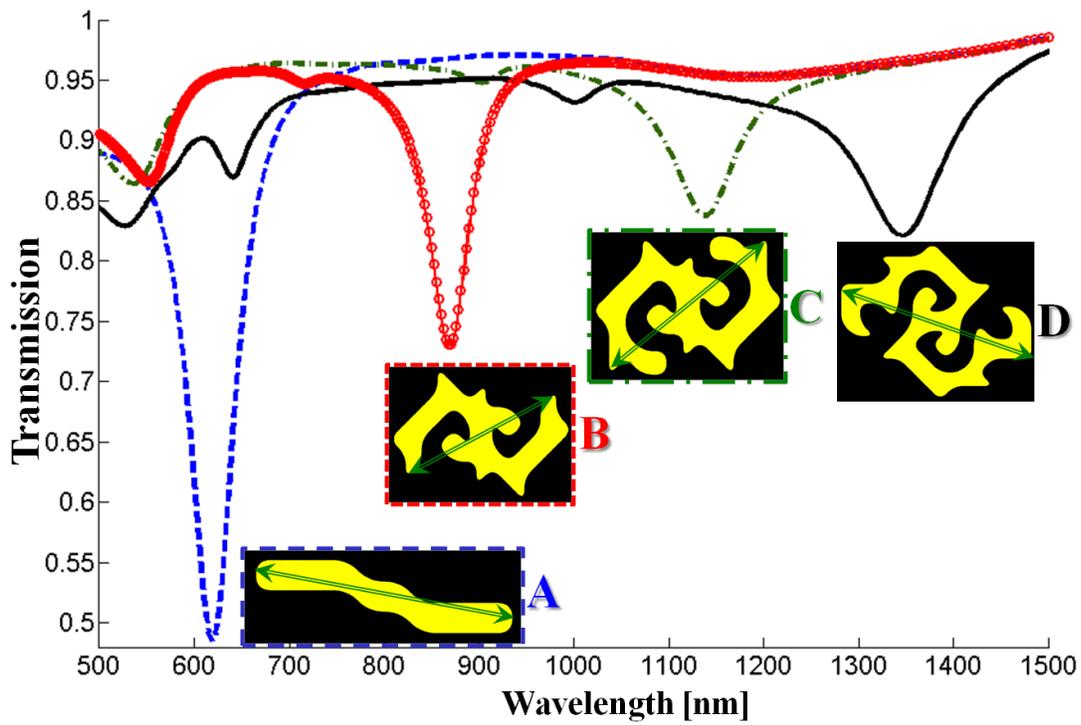

**Figure 2.**

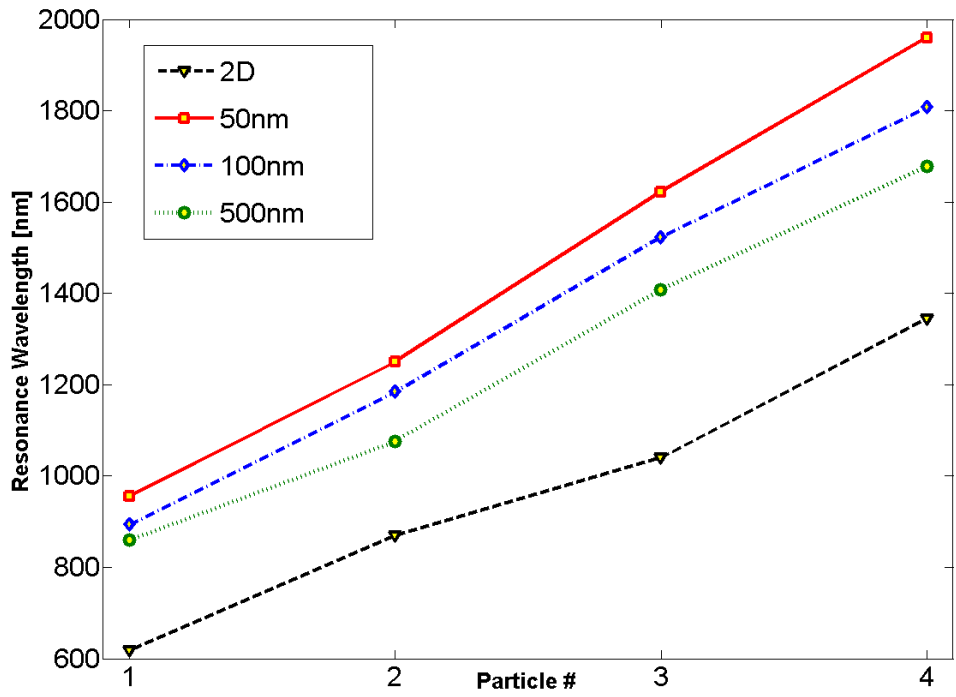

**Figure 3.**

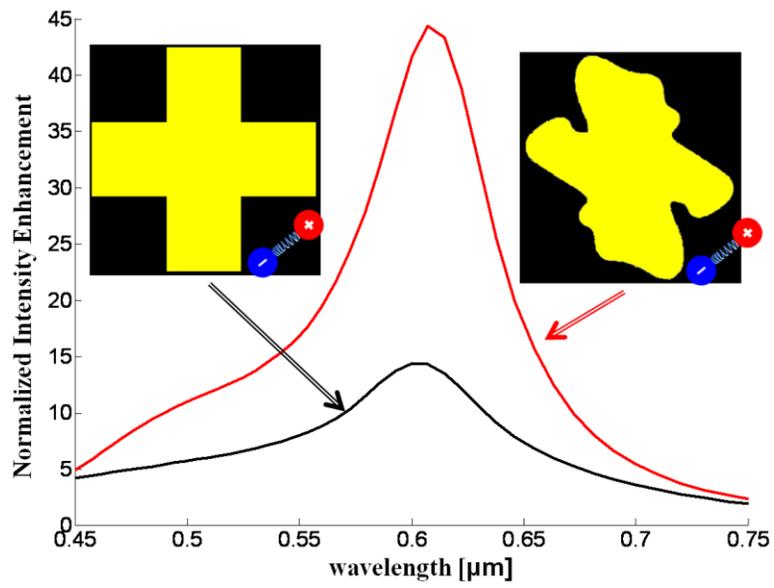

**Figure 4.**

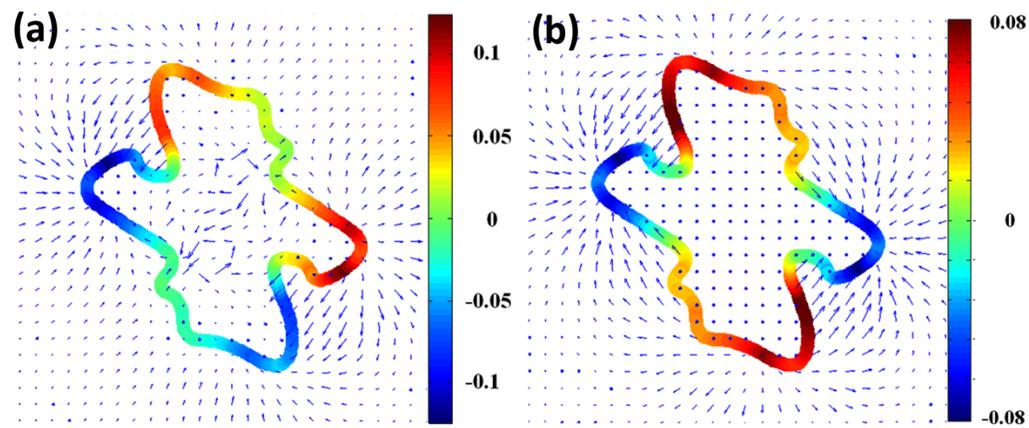